\documentclass{Interspeech2024}
\usepackage{amssymb}
\usepackage{pifont}
\usepackage{multirow}
\usepackage{subfigure}

\interspeechcameraready

\title{MFSN: Multi-perspective Fusion Search Network For Pre-training Knowledge in Speech Emotion Recognition}

\name[affiliation={1,2*}]{Haiyang}{Sun}
\name[affiliation={1*}]{Fulin}{Zhang}
\name[affiliation={1}]{Yingying}{Gao}
\name[affiliation={2}]{Zheng}{Lian}
\name[affiliation={1**}]{Shilei}{Zhang}
\name[affiliation={1**}]{Junlan}{Feng}

\address{
  $^1$China Mobile Research Institute\\
  $^2$Institute of Automation, Chinese Academy of Sciences}
\email{sunhaiyang9898@gmail.com, zhangfulin@chinamobile.com}


\begin{document}

\maketitle
  \newcommand\blfootnote[1]{%
	\begingroup
	\renewcommand\thefootnote{}\footnote{#1}%
	\addtocounter{footnote}{-1}%
	\endgroup
}
\blfootnote{*Equal contribution; **Corresponding authors}
\begin{abstract}
Speech Emotion Recognition (SER) is an important research topic in human-computer interaction. Many recent works focus on directly extracting emotional cues through pre-trained knowledge, frequently overlooking considerations of appropriateness and comprehensiveness. 
Therefore, we propose a novel framework for pre-training knowledge in SER, called Multi-perspective Fusion Search Network (MFSN). Considering comprehensiveness, we partition speech knowledge into Textual-related Emotional Content (TEC) and Speech-related Emotional Content (SEC), capturing cues from both semantic and acoustic perspectives, and we design a new architecture search space to fully leverage them. Considering appropriateness, we verify the efficacy of different modeling approaches in capturing SEC and fills the gap in current research. Experimental results on multiple datasets demonstrate the superiority of MFSN.
\end{abstract}

\section{Introduction}
\label{sec:intro}

Due to its significant contribution to human-computer interaction, Speech Emotion Recognition (SER) has received increasing attention. Researchers have been working towards endowing models with the ability to perceive and recognize emotions akin to humans. 
While some studies have incorporated various prior knowledge to guide the modeling process, the emergence of highly performant pre-trained models has led many recent works to directly leverage pre-trained knowledge for favorable outcomes. However, due to differences in training objectives, pre-training modeling methods and knowledge that excel in other downstream tasks may not perform well in SER. In this paper, we undertake a consideration and analysis from both the perspectives of appropriateness and comprehensiveness, proposing a pre-training knowledge utilization framework tailored for SER.

From the perspective of comprehensiveness, both text and speech play pivotal roles in emotion recognition \cite{kim2019dnn, wu2021emotion, lian2022smin}. However, obtaining accurate transcriptions in real-world scenarios poses challenges, and directly utilizing semantic information in speech does not enable its comprehensive exploration. Consequently, we categorize emotional cues in speech into two types: Textual-related Emotional Content (TEC), which can be extracted using Automated Speech Recognition (ASR) model to approximate text, and Speech-related Emotional Content (SEC), which can be extracted using pre-trained model to represent acoustic feature.

\begin{figure}[!ht]
	\centering
	\includegraphics[width=0.7\linewidth]{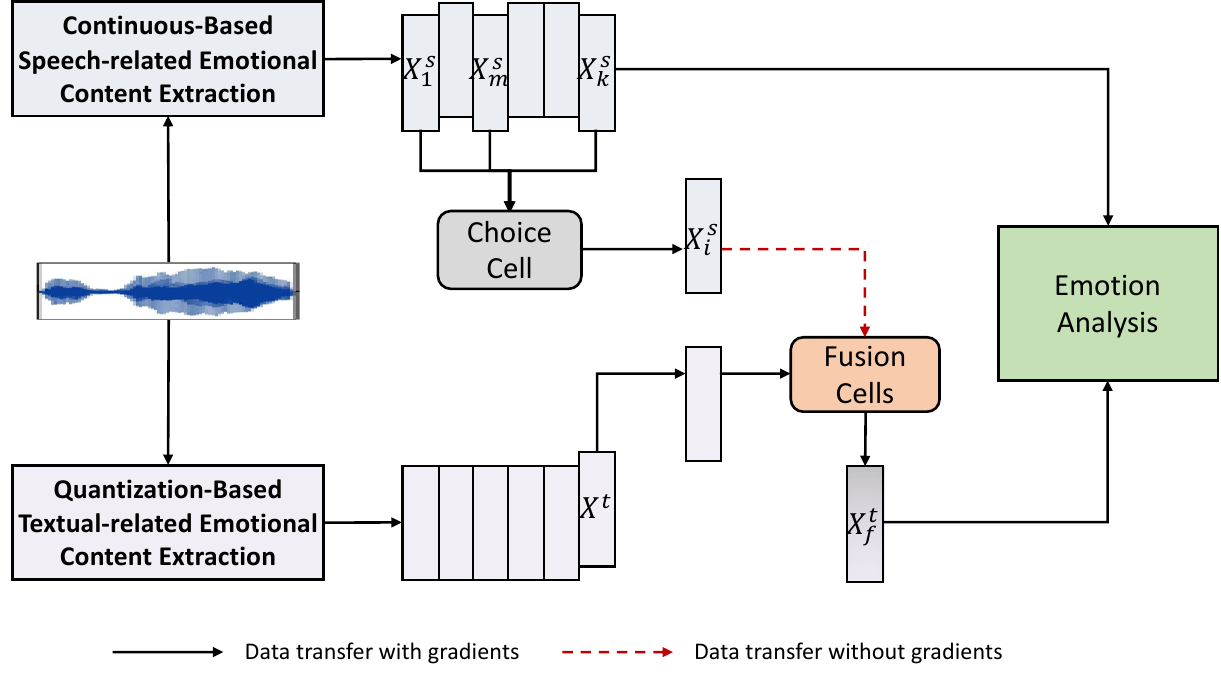}
	\caption{The overall framework of MFSN.}
	\label{MFSN}
\end{figure}

Directly leveraging both forms of information proves impractical, as transcriptions often contains emotionally charged words that may deviate from the true emotional context, potentially causing misinterpretations. For instance, someone might emphatically say, "I'm really happy right now" when they are actually angry. In such instances, individuals recalibrate their perception of emotional words based on acoustic information such as intonation or tone, or others.
We follow this principle and leverage speech feature to refine the model's understanding of TEC. 
In light of the varying significance of multi-level speech information \cite{pepino2021emotion}, the thoughtful selection of the optimal level, along with the fusion operation with TEC, becomes crucial. While exhaustive selection may yield insights, it is a time-consuming and resource-intensive process. 
Hence, we propose the use of Neural Architecture Search (NAS) for the efficient accomplishment of the aforementioned tasks.
Specifically, we design a noval search space and employ a differentiable search algorithm \cite{liu2019darts} for automated strategy design.

From the perspective of appropriateness, many contemporary approaches incorporate prior knowledge to tailor modeling methods for the target task. For instance, many existing pre-trained ASR models, grounded in inductive bias towards pronunciation units, introduce quantization operations to enhance the model's perception of text \cite{baevski2020vqwav2vec, baevski2020wav2vec2, hsu2021hubert}. Following two stages of pre-training and fine-tuning, these models can construct a constrained set of discrete units. However, there is currently no consensus on the modeling methods for SEC, often relying on pre-training speech models without due consideration \cite{zou2022speech, wang2021fine}. Given that human emotions are often dimensional and continuous \cite{plutchik1984emotions, li2020dimensional}, we follow this principle and systematically compare the performance of quantitative modeling pre-trained knowledge \cite{baevski2020wav2vec2} with continuous modeling pre-trained knowledge \cite{baevski2022data2vec} in SER.
This not only offers insights into the modeling methods for SEC but also fills current research gaps.

In conclusion, considering both comprehensiveness and appropriateness, we propose a novel framework for pre-training knowledge in SER, called Multi-perspective Fusion Search Network (MFSN). Illustrated in Figure \ref{MFSN}, MFSN employs continuous-based knowledge to capture SEC and utilizes quantization-based knowledge to capture TEC. Furthermore, it possesses the capability to automatically leverage the optimal level speech feature to refine the understanding of TEC.
Finally, both features are employed for emotion analysis.
The contributions of this paper can be summarized as follows:

\begin{itemize}
	\item We propose a novel framework for using pre-trained knowledge in SER, named MFSN. MFSN extract both TEC and SEC from appropriate perspectives. It automatically designs adjustment strategies for TEC within a newly designed search space, comprehensively exploring emotional cues.
	\item For the first time from a modeling perspective, we verify that pre-training methods based on continuous modeling are more suitable for capturing SEC than quantization. 
	\item Experimental results on multiple SER benchmark datasets demonstrate that the performance of MFSN achieves state-of-the-art levels.
\end{itemize}

\begin{table*}[t]
	\centering
	\setlength{\tabcolsep}{3pt}
	\caption{Performance comparison with existing works. Evaluation measures are UA / WA for IEMOCAP and UAR / WAR  for SAVEE.}
	\begin{tabular}{lccccc}
		\hline \hline
		\multicolumn{6}{l}{}                                                                                                                                                                                                                                                                                                                                               \\ \\[-20pt]
		\multicolumn{1}{c}{Model}               & \multicolumn{1}{c|}{\begin{tabular}[c]{@{}c@{}}IEMOCAP\\ Leave-one-session\end{tabular}} & Model              & \multicolumn{1}{c|}{\begin{tabular}[c]{@{}c@{}}IEMOCAP\\ Leave-one-speaker\end{tabular}} & Model    & SAVEE                                                                                              \\
		\multicolumn{6}{l}{}                                                                                                                                                                                                                                                                                                                                               \\ \\[-20pt] \hline
		\multicolumn{6}{l}{}                                                                                                                                                                                                                                                                                                                                               \\ \\[-20pt]
		\multicolumn{1}{c}{IS09-classification\cite{tarantino2019self}} & \multicolumn{1}{c|}{63.8 / 68.1}                                                         & CNN TF Att.pooling\cite{li2018attention} & \multicolumn{1}{c|}{68.1 / 71.8}                                                         & DCNN\cite{farooq2020impact}     & \ \ \ - \ \ \ / 82.1 \\
		\multicolumn{1}{c}{CME\cite{hang2021learning}}                 & \multicolumn{1}{c|}{73.5 / 72.7}                                                         & HNSD\cite{cao2021hierarchical}               & \multicolumn{1}{c|}{72.5 / 70.5}                                                         & TSP+INCA\cite{tuncer2021automated} & 83.4 / 84.8                                                                                        \\
		\multicolumn{1}{c}{Co-attention\cite{zou2022speech}}        & \multicolumn{1}{c|}{71.1 / 69.8}                                                         & Co-attention\cite{zou2022speech}       & \multicolumn{1}{c|}{72.7 / 71.6}                                                         & CPAC\cite{wen2022ctlmtnet}     & 83.7 / 85.6                                                                                        \\
		\multicolumn{1}{c}{Prosody2Vec\cite{qu2022disentangling}}         & \multicolumn{1}{c|}{73.3 / 72.4}                                                         & Prosody2Vec\cite{qu2022disentangling}        & \multicolumn{1}{c|}{73.9 / 72.7}                                                         & TIM-Net\cite{ye2023temporal}  & 77.3 / 79.4                                                                                        \\
		\multicolumn{6}{l}{}                                                                                                                                                                                                                                                                                                                                               \\ \\[-20pt] \hline
		\multicolumn{6}{l}{}                                                                                                                                                                                                                                                                                                                                               \\ \\[-20pt]
		\multicolumn{1}{c}{MFSN}                & \multicolumn{1}{c|}{74.0 / 71.9}                                                         & MFSN               & \multicolumn{1}{c|}{74.6 / 73.2}                                                         & MFSN     & 86.0 / 86.3                                                                                        \\
		\multicolumn{6}{l}{}                                                                                                                                                                                                                                                                                                                                               \\ \\[-20pt] \hline \hline
	\end{tabular}
	\label{sotatable}
\end{table*}

\section{Multiple-perspective Fusion Search Network}

\subsection{TEC \& SEC}
In order to comprehensively extract emotional cues from speech, we partition the emotion-related content in speech into two types: Textual-related Emotional Content (TEC) and Speech-related Emotional Content (SEC). As illustrated in Figure \ref{MFSN}, 
utilizing an ASR pre-trained model based on quantization modeling allows us to obtain TEC, referred to as $X^t$. 
Similarly, 
employing a pre-trained model based on continuous modeling with $k$ encoding layers enables us to acquire different levels of speech features, denoted as $[X_1^s, ..., X_m^s, ..., X_k^s]$, where $m=\frac{k}{2}$, $X_k^s$ is SEC. 
As the intermediate information generated during speech understanding carries different information \cite{leonardo2021emotion}, we regard $X_1^s$, $X_m^s$, and $X_k^s$ as three levels of information.
\begin{itemize}
	\item Due to the constraints of training objective, $X_k^s$ contains more target-related information.
	\item To extract $X_k^s$ from speech, the intermediate layer need to capture deeper information to fully understand content, denoted as $X_m^s$.
	\item In contrast, the output of the earlier layer, $X_1^s$, is closer to the raw speech representation.
\end{itemize}

\subsection{Search Space}
\begin{figure}[t]
	\centering
	\includegraphics[width=0.7\linewidth]{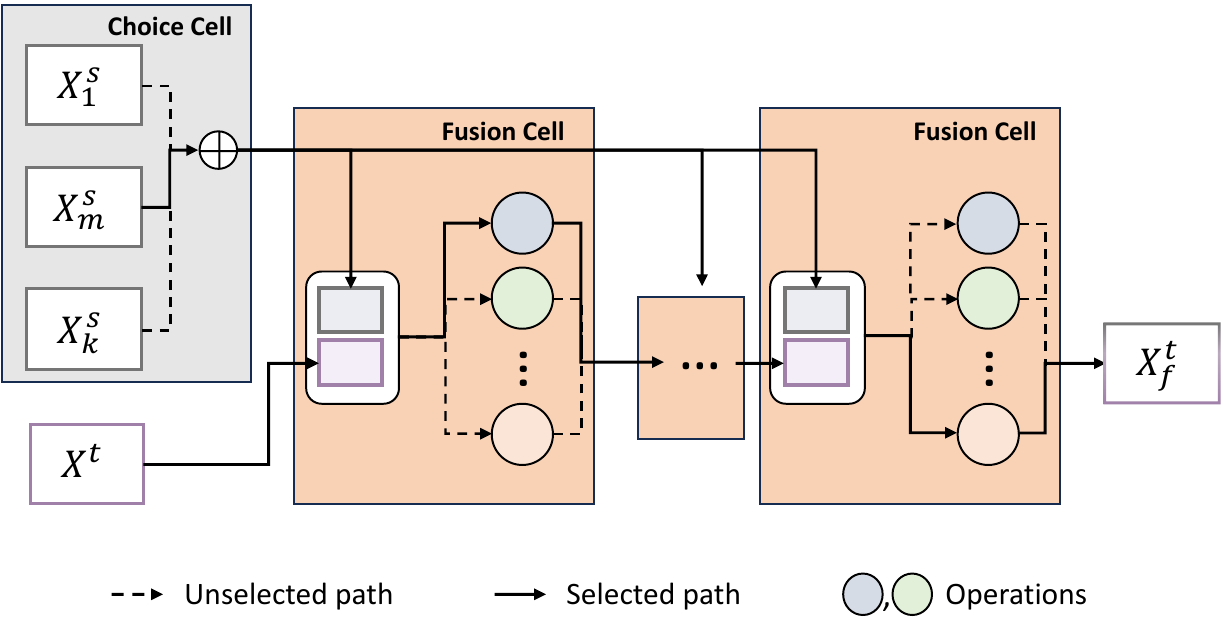}
	\caption{The search space we designed is divided into two parts: Choice Cell and Fusion Cell.}
	\label{SearchSpace}
\end{figure}
As illustrated in Figure \ref{SearchSpace}, to efficiently accomplish the selection of multi-level features and fusion operations, we design a novel search space.
Firstly, to endow the model with the capability to search different levels of speech features while avoiding potential conflicts among them, during each mini-batch training, Choice Cell randomly samples one level of SEC, denoted as $X^s_i$, where $i \in [1, m, k]$, for calculation.

Secondly, for effective adjustment of TEC and to ensure that $X^s_1$ and $X^s_m$ do not contain target-related information, we detach $X^s_i$ before feeding it into Fusion Cells along with $X^t$.

Finally, following the previous fusion work, we offer a total of eight operations in the operation pool $\mathbb{O}$ of one Fusion Cell, and each operation takes two tensor inputs and yields a output:
\begin{equation}\small
	\mathrm{Zero}(X^s_i, X^t) = 0
\end{equation}
\begin{equation}\small
	\mathrm{Sum}(X^s_i, X^t) = X^s_i + X^t
\end{equation}
\begin{equation}\small
	\mathrm{Attention}(X^s_i,X^t) = \mathrm{Softmax}(\frac{X^{s}_{i}  X^t}{\sqrt{C}})X^t
\end{equation}
\begin{equation}\small
	\begin{aligned}
		\mathrm{ConcatFC}(X^s_i,& X^t) = \\
		\mathrm{ReLU}(&\mathrm{Linear}(\mathrm{Concat}(X^s_i, X^t)))
	\end{aligned}
\end{equation}
\begin{equation}\small
	\begin{aligned}
		\mathrm{ISM}(X^s_i, X^t) = &X^s_i + \mathrm{Tanh}(\mathrm{Linear}(X^s_i) \ast H) \ast H,\\
		&\mathrm{where}\ H = \mathrm{Linear}(X^t)
	\end{aligned}
\end{equation}
Attention, ConcatFC, and ISM operations have corresponding reverse versions. For example, Attention$_r(X^s_i, X^t)$ = Attention$(X^t, X^s_i)$.  As the number of Fusion Cells, denoted as $N_f$, increases, the quantity of structures in the search space will exponentially grow to $3 * 8^{N_c}$.

\subsection{Search Algorithm}
\label{searchweight}
In the Fusion Cell, we employ a differentiable search algorithm \cite{liu2019darts} to search the fusion operations:
\begin{equation}
	X^t_f = \sum_{o\in O}{\frac{exp(\alpha_o)}{\sum_{o^{'}\in O}exp(\alpha_{o^{'}})}o(X^s_i, X^t)}
\end{equation}
where $\alpha \in \mathbb{R}^{|\mathbb{O}|}$ is a trainable weight that measures the importance of different operations, and $|\mathbb{O}|$ denotes the number of operations in $\mathbb{O}$. With the assistance of NAS, it is possible to avoid exhaustive training of all $3 * 8^{N_c}$ structures and find the optimal adjustment strategy through a single training process.  Both $X^s_k$ and $X^t_f$ are subsequently used for emotional analysis.

\section{Different modeling methods}
To choose the most appropriate knowledge for extracting SEC, we conduct a comprehensive analysis of various pre-training modeling methods. The current modeling methods can be broadly categorized into two types: quantization-based reconstruction \cite{baevski2020vqwav2vec, baevski2020wav2vec2, hsu2021hubert} and continuous frame-based reconstruction \cite{liu2021tera, baevski2022data2vec}. Both approaches aim to understand the content of speech by reconstructing the masked portions. The key difference between these approaches lies in their representation space of speech frames. Formally, we represent the speech as $X = [x_1, ..., x_T]$, where $T$ is the number of frames. Given a set of indices $Z = [z_1, ..., z_{T'}]$ corresponding to the frames that need to be masked, where $0 \le T' \leq  T$, the frames at the indices in $Z$ are replaced with a trainable code word $c$, resulting in $\tilde{X} = [..., x_t, ..., c, ...]$, where $0 \le t < T$. The model then encodes $\tilde{X}$ to obtain its output $\hat{X} = [\hat{x}_1, ..., \hat{x}_T]$.

Quantization-based reconstruction methods discretize the representation space of speech frames by clustering them into $N$ centroids, denoted as $Q = [q_1, ..., q_N]$. For each frame in $X[Z]$, the corresponding cluster center is used as the label, yielding a label vector $Y = [q_i, ..., q_j]$, where $0 \le  i, j \le N$. The objective of the model is to minimize the discrepancy between $\hat{X}[Z]$ and $Y$. On the other hand, continuous frame reconstruction method directly use the original frames at the masked positions as the labels, i.e., $Y=X[Z]$.

\begin{figure}[t]
	\centering
	\includegraphics[width=0.7\linewidth]{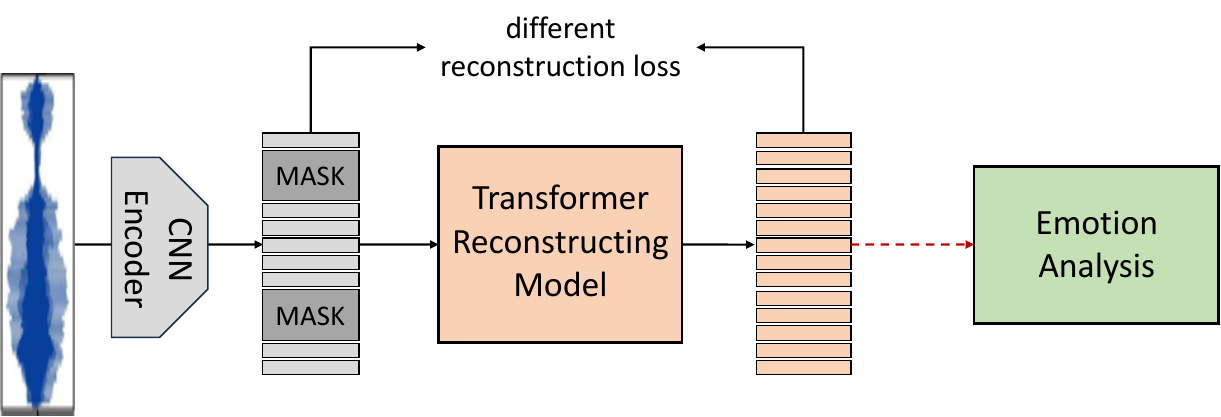}
	\caption{Unified training framework for different modeling.}
	\label{Toy}
\end{figure}

\begin{table}[t]
	\centering
	\caption{Performance comparison with works using text feature.}
	\begin{tabular}{ccc|c}
		\hline \hline
		\begin{tabular}[c]{@{}c@{}}Speech\\ Feature\end{tabular} & \begin{tabular}[c]{@{}c@{}}Text\\ Feature\end{tabular} & Model & \begin{tabular}[c]{@{}c@{}}IEMOCAP\\ Leave-one-session\end{tabular} \\ \hline
		MFCC                                              & -                                                     & BLSTM\cite{santoso2022speech} & 61.1 / 64.3                                                         \\
		-                                                        & ASR+BERT                                               & BLSTM\cite{santoso2022speech} & 71.8 / 71.9                                                         \\
		MFCC                                              & ASR+BERT                                               & SAWC\cite{santoso2022speech}  & 76.8 / 76.6                                                         \\
		SEC                                                      & TEC                                                    & MSFN  & 74.0 / 71.9                                                         \\ \hline \hline
	\end{tabular}
	\label{Textfeature}
\end{table}

\section{Datasets}
Interactive Emotional Dyadic Motion Capture (IEMOCAP) \cite{busso2008iemocap} dataset is a widely adopted benchmark for evaluating emotion recognition models. Following prior research, we focus on four emotions: angry, sad, happy, and neutral. To ensure a robust evaluation of our method, we adopted the 5-fold leave-one-session-out and 10-fold leave-one-speaker-out validation strategies to generate results.

Surrey Audio-Visual Expressed Emotion (SAVEE) \cite{savee} dataset consists of recordings from 4 male actors in 7 different emotions. Following prior research, we adopted the 10-fold cross-validation strategies to generate results.

\section{Experimental Setup \& Results}

\subsection{Main setup}
In this paper, we adopt Co-attention \cite{zou2022speech} as our Emotion Analysis module. Thus, in data preprocessing, in alignment with it, we segmented each speech into several 3-second-long segments and extracted the corresponding spectrogram and MFCC features for these segments. The final prediction for each speech was determined by aggregating the predictions from all its segmented parts.

In MFSN, we utilize pre-trained knowledge based on continuous modeling from Data2vec \cite{baevski2022data2vec} to extract SEC from speech. For TEC, we retain prior knowledge of pronunciation units and capture it using fine-tuned Wav2vec2 \cite{baevski2020wav2vec2}. Additionally, we employed an SGD optimizer with a learning rate of 5 to optimize the operation weight $\alpha$, as described in Section \ref{searchweight}. For the convenience of analysis, we employ a single Fusion Cell. When dealing with MFSN dual-stream inputs, we double the dimension of the encoding hidden features for both spectrograms and MFCCs. These features are evenly distributed to each stream for co-attention calculations.

\subsection{Comparison with existing works} 
Following prior works, we conducted performance tests of MFSN on both the IEMOCAP and SAVEE datasets. As depicted in Table \ref{sotatable}, in our experiments on the IEMOCAP dataset, MFSN attains the highest Unweighted Accuracy (UA) score in the Leave-one-session cross-validation strategy, demonstrating competitive Weighted Accuracy (WA) scores even in the presence of class imbalance. Furthermore, within the Leave-one-speaker cross-validation strategy, MFSN consistently achieves the top UA and WA scores. When applied to the broader range of categories in the SAVEE dataset, MFSN exhibits a significant advantage, surpassing the sota in the 10-fold cross-validation.

However, as shown in Table \ref{Textfeature}, there is still a gap between TEC and BERT. Although MFSN surpass the performance of ASR+BERT, it still lags behind the SAWC \cite{santoso2022speech}.

\subsection{Comparison between different modeling methods}
\subsubsection{Unified model configuration}

\begin{table}[t]\small
	\centering
	\caption{Speech encoder architecture.}
	\begin{tabular}{ccc}
		\hline \hline
		\multirow{3}{*}{CNN Encoder} & strides         & 5, 2, 2, 2, 2, 2, 2  \\
		& kernel width    & 10, 3, 3, 3, 3, 2, 2 \\
		& channel         & 512                  \\ \hline
		\multirow{4}{*}{Transformer} & layer           & 4                    \\
		& embedding dim.  & 512                  \\
		&  FFN dim.  & 2048                 \\
		& attention heads & 8                    \\ \hline \hline
	\end{tabular}
	\label{Table2}
\end{table}

\begin{table*}[h]\small
	\centering
	\setlength{\tabcolsep}{3pt}
	\caption{The results of emotion analysis using knowledge learned from different modeling approaches.}
	\begin{tabular}{cccccccccccccc}
		\hline \hline
		\multicolumn{1}{l}{}                                                           & \multicolumn{1}{l}{}                                                         & \multicolumn{1}{l}{} & \multicolumn{1}{l}{} & \multicolumn{1}{l}{} & \multicolumn{1}{l}{} & \multicolumn{1}{l}{}        & \multicolumn{1}{l}{} & \multicolumn{1}{l}{} & \multicolumn{1}{l}{} & \multicolumn{1}{l}{} & \multicolumn{1}{l}{}        & \multicolumn{1}{l}{} & \multicolumn{1}{l}{} \\ \\[-20pt]
		& \multicolumn{1}{c|}{}                                                        & \multicolumn{10}{c|}{IEMOCAP}                                                                                                                                                                                                                     & \multicolumn{2}{c}{\multirow{2}{*}{SAVEE}}  \\ \cline{4-11}
		& \multicolumn{1}{c|}{}                                                        & \multicolumn{5}{c}{Leave-one-sessiong}                                                                                  & \multicolumn{5}{c|}{Leave-one-speaker}                                                                                  & \multicolumn{2}{c}{}                        \\
		\multicolumn{1}{l}{}                                                           & \multicolumn{1}{l}{}                                                         & \multicolumn{1}{l}{} & \multicolumn{1}{l}{} & \multicolumn{1}{l}{} & \multicolumn{1}{l}{} & \multicolumn{1}{l}{}        & \multicolumn{1}{l}{} & \multicolumn{1}{l}{} & \multicolumn{1}{l}{} & \multicolumn{1}{l}{} & \multicolumn{1}{l}{}        & \multicolumn{1}{l}{} & \multicolumn{1}{l}{} \\ \\[-20pt] \hline
		\multicolumn{1}{l}{}                                                           & \multicolumn{1}{l}{}                                                         & \multicolumn{1}{l}{} & \multicolumn{1}{l}{} & \multicolumn{1}{l}{} & \multicolumn{1}{l}{} & \multicolumn{1}{l}{}        & \multicolumn{1}{l}{} & \multicolumn{1}{l}{} & \multicolumn{1}{l}{} & \multicolumn{1}{l}{} & \multicolumn{1}{l}{}        & \multicolumn{1}{l}{} & \multicolumn{1}{l}{} \\ \\[-20pt]
		\multicolumn{1}{c|}{\begin{tabular}[c]{@{}c@{}}Modeling\\ Method\end{tabular}} & \multicolumn{1}{c|}{\begin{tabular}[c]{@{}c@{}}Model\\ Setting\end{tabular}} & V                    & A                    & D                    & UA(\%)               & \multicolumn{1}{c|}{WA(\%)} & V                    & A                    & D                    & UA(\%)               & \multicolumn{1}{c|}{WA(\%)} & UA(\%)               & WA(\%)               \\
		\multicolumn{1}{l}{}                                                           & \multicolumn{1}{l}{}                                                         & \multicolumn{1}{l}{} & \multicolumn{1}{l}{} & \multicolumn{1}{l}{} & \multicolumn{1}{l}{} & \multicolumn{1}{l}{}        & \multicolumn{1}{l}{} & \multicolumn{1}{l}{} & \multicolumn{1}{l}{} & \multicolumn{1}{l}{} & \multicolumn{1}{l}{}        & \multicolumn{1}{l}{} & \multicolumn{1}{l}{} \\ \\[-20pt] \hline
		\multicolumn{1}{l}{}                                                           & \multicolumn{1}{l}{}                                                         & \multicolumn{1}{l}{} & \multicolumn{1}{l}{} & \multicolumn{1}{l}{} & \multicolumn{1}{l}{} & \multicolumn{1}{l}{}        & \multicolumn{1}{l}{} & \multicolumn{1}{l}{} & \multicolumn{1}{l}{} & \multicolumn{1}{l}{} & \multicolumn{1}{l}{}        & \multicolumn{1}{l}{} & \multicolumn{1}{l}{} \\ \\[-20pt]
		\multicolumn{1}{c|}{\multirow{4}{*}{Quantization}}                             & \multicolumn{1}{c|}{$B$=2 $W$=2 $N$=4}                                       & 1.02                 & 0.54                 & 0.71                 & 60.3                 & \multicolumn{1}{c|}{59.6}   & 1.07                 & 0.60                 & 0.71                 & 62.4                 & \multicolumn{1}{c|}{61.9}   & 37.1                 & 43.5                 \\
		\multicolumn{1}{c|}{}                                                          & \multicolumn{1}{c|}{$B$=2 $W$=8 $N$=64}                                      & 1.13                 & 0.49                 & 0.68                 & 60.6                 & \multicolumn{1}{c|}{60.3}   & 1.06                 & 0.44                 & 0.64                 & 63.8                 & \multicolumn{1}{c|}{63.3}   & 42.9                 & 47.9                 \\
				\multicolumn{1}{c|}{}                                                          & \multicolumn{1}{c|}{$B$=2 $W$=12 $N$=144}                                    & 1.01                 & 0.47                 & 0.65                 & 60.6                 & \multicolumn{1}{c|}{60.2}   & 0.95                 & 0.40                 & 0.57                 & 63.2                 & \multicolumn{1}{c|}{62.3}   & 45.8                 & 49.2                 \\
		\multicolumn{1}{c|}{}                                                          & \multicolumn{1}{c|}{$B$=4 $W$=8 $N$=4096}                                    & 0.72                 & 0.28                 & 0.46                 & 60.3                 & \multicolumn{1}{c|}{60.2}   & 0.70                 & 0.28                 & 0.45                 & 62.6                 & \multicolumn{1}{c|}{62.6}   & 47.3                 & 51.3                 \\
		\multicolumn{1}{l}{}                                                           & \multicolumn{1}{l}{}                                                         & \multicolumn{1}{l}{} & \multicolumn{1}{l}{} & \multicolumn{1}{l}{} & \multicolumn{1}{l}{} & \multicolumn{1}{l}{}        & \multicolumn{1}{l}{} & \multicolumn{1}{l}{} & \multicolumn{1}{l}{} & \multicolumn{1}{l}{} & \multicolumn{1}{l}{}        & \multicolumn{1}{l}{} & \multicolumn{1}{l}{} \\ \\[-20pt] \hline
		\multicolumn{1}{l}{}                                                           & \multicolumn{1}{l}{}                                                         & \multicolumn{1}{l}{} & \multicolumn{1}{l}{} & \multicolumn{1}{l}{} & \multicolumn{1}{l}{} & \multicolumn{1}{l}{}        & \multicolumn{1}{l}{} & \multicolumn{1}{l}{} & \multicolumn{1}{l}{} & \multicolumn{1}{l}{} & \multicolumn{1}{l}{}        & \multicolumn{1}{l}{} & \multicolumn{1}{l}{} \\ \\[-20pt]
				\multicolumn{1}{c|}{\multirow{2}{*}{Continuous}}                               & \multicolumn{1}{c|}{$L$=1}                                                   & 0.71                 & 0.28                 & 0.45                 & 60.5                 & \multicolumn{1}{c|}{60.3}   & 0.68                 & 0.28                 & 0.44                 & 62.5                 & \multicolumn{1}{c|}{62.6}   & 59.1                 & 61.3                 \\
		
		\multicolumn{1}{c|}{}                                                          & \multicolumn{1}{c|}{$L$=4}                                                   & 0.70                 & 0.28                 & 0.45                 & 60.6                 & \multicolumn{1}{c|}{60.6}   & 0.68                 & 0.28                 & 0.44                 & 62.8                 & \multicolumn{1}{c|}{62.5}   & 60.6                 & 63.1                 \\
		\multicolumn{1}{l}{}                                                           & \multicolumn{1}{l}{}                                                         & \multicolumn{1}{l}{} & \multicolumn{1}{l}{} & \multicolumn{1}{l}{} & \multicolumn{1}{l}{} & \multicolumn{1}{l}{}        & \multicolumn{1}{l}{} & \multicolumn{1}{l}{} & \multicolumn{1}{l}{} & \multicolumn{1}{l}{} & \multicolumn{1}{l}{}        & \multicolumn{1}{l}{} & \multicolumn{1}{l}{} \\ \\[-20pt] \hline \hline
	\end{tabular}
	\label{Tabletest}
\end{table*}

As shown in Figure \ref{Toy}, to mitigate the impact caused by variations in engineering implementations, we employ a unified framework for different modeling methods. The specific parameters of the network are shown in Table \ref{Table2}. In the quantization modeling method, we imitate Wav2vec2 \cite{baevski2020wav2vec2} and assign $B$ codebooks to all frame features, each containing $W$ codewords, resulting in a total of $N = W^B$ clustering centers.  The label vector $Y$ is composed of the clustering centers corresponding to $X[Z]$ and the sampled negative examples. 
In the continuous modeling method, we imitate Data2vec \cite{baevski2022data2vec} and use the mean value of the outputs of the last $L$ encoder layers as the $Y$.
They were trained separately using BCE loss and MSE loss for reconstruction, with a data mask rate set at 65\%.
\subsubsection{Results}
\label{section3}

To fully compare the capabilities of various modeling methods in capturing SEC, we not only test the models' capability in discrete emotion recognition but also in dimensional emotion analysis. In Table \ref{Tabletest}, we report the MSE metrics of Valence (V), Activation (A), Dominance (D), and report UA,  WA.

As demonstrated by the performance in V, A, and D, in dimensional emotion analysis, knowledge acquired through quantization modeling exhibits notable limitations. The performance significantly improves as the number of quantization units, $N$, increases. Only when $N$ is large enough can this method approach the performance of continuous modeling. In the continuous modeling, the variation of the encoding layers $L$ as reconstructed labels does not noticeably affect performance.

For discrete emotion recognition, in the four-class IEMOCAP task, the quantization modeling shows comparable results to the continuous modeling. However, in more complex emotional categorizations, such as the seven-class SAVEE task, quantization modeling leads to a noticeable performance decline. Similarly, inadequate $N$ numbers restrict model performance, while various settings of continuous modeling exhibit relatively consistent performance.

The above experimental results demonstrate that, whether in discrete emotion recognition or dimensional emotion analysis, the pre-trained knowledge derived from continuous modeling proves to be a superior choice for extracting SEC.

\begin{table}[]
	\centering
	\caption{Performance comparison between different model configurations. Evaluation measures are UA(\%) / WA(\%).}
	\begin{tabular}{cccc}
		\hline \hline
		&             &                          &                                                                     \\ \\[-20pt]
		SEC Encoder & TEC Encoder & \multicolumn{1}{c|}{NAS} & \begin{tabular}[c]{@{}c@{}}IEMOCAP\\ Leave-one-session\end{tabular} \\
		&             &                          &                                                                     \\ \\[-20pt] \hline
		&             &                          &                                                                     \\ \\[-20pt]
		Wav2vec2    & -           & \multicolumn{1}{c|}{\ding{53}}    & 70.6 / 69.6                                                         \\
		Data2vec    & -           & \multicolumn{1}{c|}{\ding{53}}    & 72.9 / 71.8                                                         \\
		Data2vec    & Wav2vec2    & \multicolumn{1}{c|}{\ding{53}}    & 73.1 / 71.7                                                         \\
		Data2vec    & Wav2vec2    & \multicolumn{1}{c|}{\checkmark}    & 74.0 / 71.9                                                         \\
		&             &                          &                                                                     \\ \\[-20pt] \hline \hline
	\end{tabular}
	\label{ablation}
\end{table}

\subsection{Comparison with different configurations}
We further analyze the importance of each module in MFSN by evaluating various model configurations using the Leave-one-session strategy on IEMOCAP. 
As shown in Table \ref{ablation}, consistent with the findings in Section \ref{section3}, the pre-trained knowledge based on continuous modeling in Data2vec better captures SEC.
However, simply incorporating TEC without adjustment, resulting in no performance improvement and even a slight decrease in the WA metric. Through the adjustment strategy search by MFSN, the model leverages SEC to mitigate bias in TEC, leading to a further improvement in model performance.

\subsection{Visualization of Search Results}
\begin{figure}[t]
	\centering
	\includegraphics[width=0.7\linewidth]{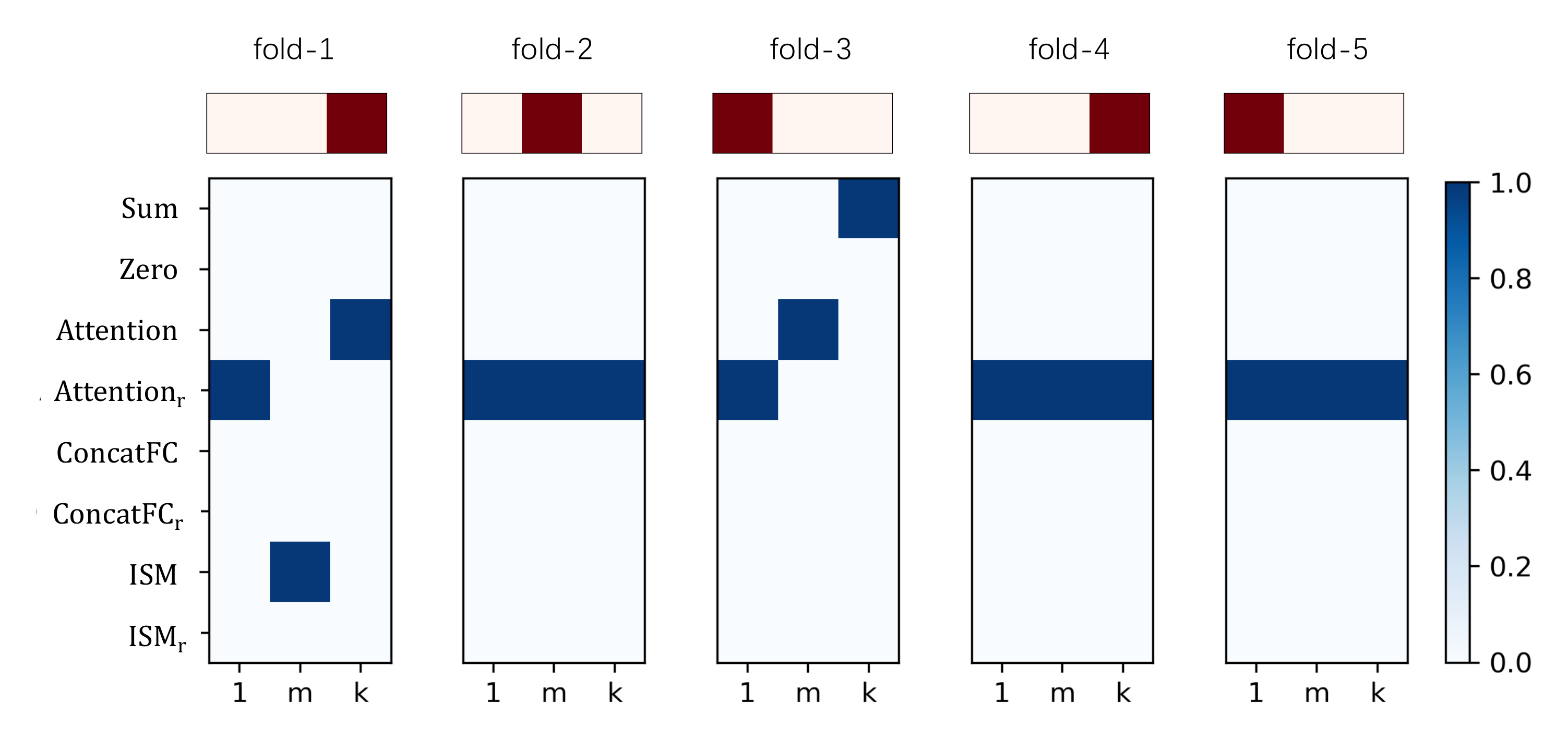}
	\caption{The visualization of adjustment strategy search results for the Leave-one-session strategy. Here, $1$, $m$, and $k$ represent three levels of features. Red color indicates the best path. }
	\label{T}
\end{figure}

As shown in Figure \ref{T}, in various data partition scenarios, the Choice Cell and Fusion Cell automatically select optimal speech features and fusion operations to adjust TEC. To further demonstrate the performance of MFSN on imbalanced datasets, we present the performance confusion matrix of MFSN compared to the Co-attention method using only Wav2vec2 as the TEC Encoder, as shown in Figure 4.

\begin{figure}[t]
	\centering
	\subfigure[Co-attention \cite{zou2022speech}]{
		\includegraphics[width=0.35\linewidth]{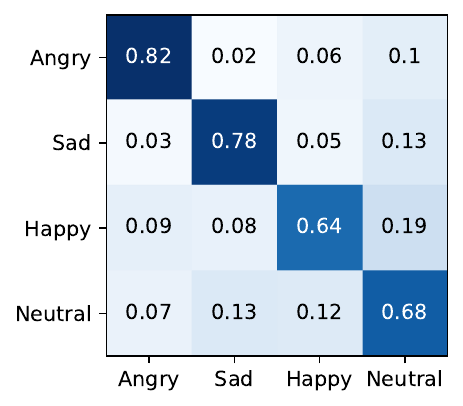}
	}
	\subfigure[MFSN]{
		\label{Modeltokenmixer}
		\includegraphics[width=0.35\linewidth]{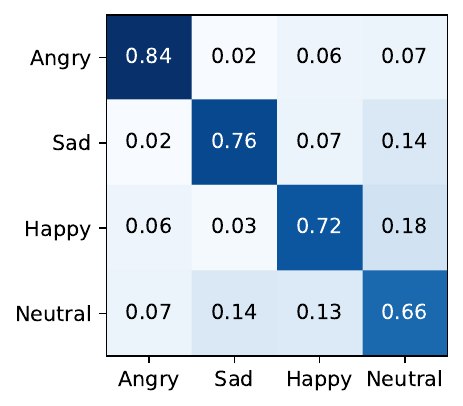}
	}
	\caption{Visualization of confusion matrices.}
	\label{T2}
\end{figure}

\section{Conclusions}
In this paper, we propose a novel framework for pre-training knowledge in SER, called MFSN. Operating within a novel search space, it comprehensively captures emotional cues, encompassing both Speech-related Emotional Content and Textual-related Emotional Content. Through in-depth comparisons across various pre-train modeling methods, MFSN leverages more appropriate knowledge to extract SEC. Experimental results show that MFSN outperforms existing methods.

\bibliographystyle{IEEEtran}
\bibliography{mybib}

\end{document}